\documentclass[aps,prd,showpacs,amsmath,amssymb]{revtex4}

\begin{document}

\title{Generalized unparticles, zeros of the Green function, and momentum
space topology of the lattice model with overlap fermions}

\author{M.A.Zubkov}
 \email{zubkov@itep.ru}
\affiliation{ITEP, B.Cheremushkinskaya 25, Moscow, 117259, Russia}

\today


\begin{abstract}
The definition of topological invariants $\tilde{\cal N}_4, \tilde{\cal N}_5$
suggested in \cite{VZ2012} is extended to the case, when there are zeros and
poles of the Green function in momentum space.  It is shown how to extend the
index theorem suggested in \cite{VZ2012} to this case. The non - analytical
exceptional points of the Green function appear in the intermediate vacuum,
which exists at the transition line between the massive vacua with different
values of topological invariants.   Their number is related to the jump
$\Delta\tilde{\cal N}_4$ across the transition. The given construction is
illustrated by momentum space topology of the lattice model with overlap
fermions. In the vicinities of the given points the fermion excitations appear
that cannot be considered as usual fermion particles. We, therefore, feel this
appropriate to call them generalized unparticles. This notion is, in general
case different from the Georgi's unparticle. However, in the case of lattice
overlap fermions the propagator of such excitations is indeed that of the
fermionic unparticle suggested in \cite{fermion_unparticle}.
\end{abstract}


\maketitle

\newtheorem{theorem}{Theorem}[section]
\newtheorem{hypothesis}[theorem]{Hypothesis}
\newtheorem{lemma}{Lemma}[section]
\newtheorem{corollary}[lemma]{Corollary}
\newtheorem{proposition}[lemma]{Proposition}
\newtheorem{claim}[lemma]{Claim}

\newtheorem{definition}[lemma]{Definition}
\newtheorem{assumption}{Assumption}

\newtheorem{remark}[lemma]{Remark}
\newtheorem{example}[lemma]{Example}
\newtheorem{problem}[lemma]{Problem}
\newtheorem{exercise}[lemma]{Exercise}



\newcommand{\br}{{\bf r}}
\newcommand{\bu}{{\bf \delta}}
\newcommand{\bk}{{\bf k}}
\newcommand{\bq}{{\bf q}}
\def\({\left(}
\def\){\right)}
\def\[{\left[}
\def\]{\right]}

\section{Introduction}

Recently, in condensed matter physics there was some discussion of the possible
role of unusual exceptional points of the Green function. Namely, the zeros of
the Green function may appear as a result of unusual renormalization (see, for
example, Eq. (9) in \cite{Volovik2007}, where zeroes in the Green's function
instead of pole emerge due to strong interactions in cuprates in the so-called
pseudogap state \cite{KuSa}). Also non - analytical exceptional points may
appear. For example,  in Eq.(6) of \cite{Volovik2011} such points emerge from
poles due to interactions. In \cite{EssinGurarie2011} and \cite{Gurarie2011}
the role of zeroes in Green function in topological insulators is discussed.
The formation of zeros of the Green function was also discussed in
\cite{SilaevVolovik2012}. On the other hand the role of such exceptional points
is not known in details.

In high energy physics the non - analytical exceptional points of the Green
function were discussed as well. For example, in \cite{unparticle} the
elementary excitations of the quantum fields were considered that do not look
like particles. The corresponding operators have unusual scaling dimensions and
belong to  a hidden scale - invariant sector of the field theory. These objects
were called unparticles because their properties differ from that of the usual
particle - like excitations. In particular, the propagators of unparticles may
have non - analytical exceptional points at $p = 0$ (see \cite{unparticle}).
The form of the propagator for the fermion unparticle was suggested in
\cite{fermion_unparticle}. Further also another form of the  fermionic
propagator for unparticles was suggested in \cite{fermion_unparticle2}. The
possible relation of the unparticle physics to TeV - phenomenology has been
discussed actively over the last years (see, for example,
\cite{unparticle_phenomenology}, and references therein).

In the present paper we consider the case, when the non - analytical
exceptional point of general form appears in momentum space of the relativistic
fermionic system. In the vicinity of such a point the excitations appear that
also do not look like ordinary particles. We call them generalized unparticles.
It is worth mentioning that this notion as it is used here is more wide than
the notion of usual fermionic unparticles. Namely, the generalized unparticle
has the propagator with non - analyticity of general form while in
\cite{fermion_unparticle, fermion_unparticle2} the particular forms of the
propagator were derived in accordance with the given scaling dimension $d_U$.
Nevertheless, the propagator of the generalized unparticles that appear in the
system of free overlap fermions coincides with that of suggested in
\cite{fermion_unparticle} for $d_U = 2$.

In the present paper we generalize the approach of \cite{VZ2012,Volovik2010},
where momentum space topology of the relativistic gapped models was considered
(see also
\cite{NielsenNinomiya1981,So1985,IshikawaMatsuyama1986,Horava2005,Creutz2008,Kaplan2011,Kaplan1992,Golterman1993,Volovik2003}
and \cite{Volovik2011,HasanKane2010,Xiao-LiangQi2011,Wen2012}). In
\cite{VZ2012} the topological invariants were constructed for the gapped
fermionic systems. These topological invariants are related to the number of
massless fermions existing at the phase transitions between different gapped
vacua. Here we generalize the topological invariants $\tilde{\cal N}_4$,
$\tilde{\cal N}_5$ to the case when there are zeros and poles of the Green
function in momentum space. In their presence at the phase transition between
the vacua of the fermionic system with different values of $\tilde{\cal N}_4$
the generalized unparticles appear. Their number as well as the number of
massless particles at the transition is related to the jumps of $\tilde{\cal
N}_4$ and $\tilde{\cal N}_5$.

The momentum space topology of the fermionic system with zeros of the Green
function and generalized unparticles is illustrated by the example of the
lattice model with overlap fermions \cite{Creutz2011,Overlap,Shrock}. Table
\ref{table4} demonstrates the values of the topological invariants in different
ranges of mass parameters $m$, $m_0$ in which the vacuum represents different
phases of topological insulator. There are critical values of mass parameter
$m_0 = 2$, $4$, $6$ and $8$, at which for $m \ne 0$ the topological quantum
phase transitions occur between the insulators with different values of
$\tilde{\cal N}_4$ and $\tilde{\cal N}_5$ (more on topological phase
transitions, at which the topological charge of the vacuum changes while the
symmetry does not, see \cite{Volovik2007}). At these values of $m_0$ the vacuum
states contain the generalized unparticles. At the same time both in these
intermediate states and in the insulator states the zeros of the Green function
are present in momentum space.  There is an analogue of the index theorem
presented in \cite{VZ2012},  which relates the number of the mentioned
generalized unparticles with the jump $\Delta\tilde{\cal N}_4$ of the
topological invariant across the transition. The total number of topologically
protected unparticles at the transition point is $n^u_F=\Delta \tilde{\cal N}_4
$, and according to Table \ref{table4} the states with critical $m_0 = 2$, $4$,
$6$ and $8$  have correspondingly 1, 4, 6, 4, 1 generalized unparticle species.

For the values of $m_0 \ne 0,2,4,6,8$ at the phase transition between the
states with different signs of $m$ there are no unparticles but massless
fermion excitations appear. The number of topologically protected massless
fermions $n_f^0$ is related to the jump of $\tilde{\cal N}_5$: $n_f^0 = \Delta
\tilde{\cal N}_5/2$. However, the observed number $n_f$ of massless fermions in
these states is larger than $n_f^0$ except for the case $2>m_0>0$ that is used
in conventional discretization. This means that for $m >2$ turning on gauge
fields we may decrease the number of massless fermions in the intermediate
states without a phase transition.

For the value $m = 0$ quantum phase transitions are observed at $m_0 =
2,4,6,8$. At these transitions unparticles appear with the propagator equal (up
to the scaling factor) to that of given in \cite{fermion_unparticle} for $d_U =
2$. The number of such unparticles is equal to the jump of $\tilde{\cal N}_4$
across the transition.

All invariants mentioned above are expressed in terms of the Green's function,
and are applicable to the interacting systems as well \cite{Volovik2003}, so
that the relation between the number of unparticles/massless excitations and
the jumps in the topological charges remains valid when the interactions
between fermions are turned on.

The main subject of the present paper is the generalization of topological invariants and index theorem to the case when the fermion Green function may contain zeros and nonanalytical exceptional points. A very personal 
expectation is that models with such unusual properties may be relevant for the description of the new TeV scale physics expected to appear at the LHC. Say, a unified field theory that works somewhere above TeV, may have the Green functions with zeros, our world may live at the position of the phase transition within a model of this kind. The massless fermions and the generalized unparticles appear at the position of this transition. Their numbers are related to the jumps of the topological invariants across the transition. Originally, the role of the lattice model with overlap fermions was to illustrate this construction. However, the consideration of this model at “unphysical” values of $m_0$ is interesting itself. The structure of the lattice model including the properties of various lattice artifacts is to be investigated in order to extract physical information from the given lattice model in a proper way. In addition, the model with overlap fermions may indeed be used as a formulation of the theory that contains the fermionic unparticles with the scaling dimension $d_U=2$. Also, the constructions presented in the present paper, in principle, may be applied in a some way to the investigation of the Aoki phase. However, this would require additional efforts and is out of the scope of the present paper. 

The paper is organized as follows. In Section \ref{Sect5D4D} the constructions
presented in \cite{VZ2012} are reviewed. In Section \ref{SectIndTheor} we
extend the definitions of $\tilde{\cal N}_4, \tilde{\cal N}_5$ given in
\cite{VZ2012} to the case, when zeros and poles of the Green function are
present. Also we represent in this section the generalized index theorem. In
Section \ref{lattice} we consider momentum space topology of the lattice system
with overlap fermions. In Section \ref{Conclusions} we end with the
conclusions.

\section{Topological invariants for the case, when there are no zeros and poles of the Green function}
\label{Sect5D4D}

\subsection{Green functions}

As well as in \cite{VZ2012} here and below we consider the fermion systems with
the Green function in Euclidean momentum space of the form
 \begin{equation}
  {\cal G} = \frac{Z[p^2]}{g^i[p] \gamma^i -
im[p]}, \quad i = 1,2,3,4 \label{G_}
\end{equation}
Here $Z(p^2)$ is the wave function renormalization function, while $m(p^2)$ is
the effective mass term, $g_a[p]$ are real functions. $Z$, $g$, and $m$ may be
though of as the diagonal matrices if several flavors of Dirac fermions are
present. The lattice model with noninteracting overlap fermions has the Green
function of the form Eq. (\ref{G_}) as it will be seen in Section
\ref{lattice}. Also the model with the overlap fermions coupled to Gauge fields
may have the Green function of the same form \cite{Overlap}. In general case
(and, in particular, for the overlap fermions) the Green function may have both
poles and zeros. However, in the present Section we consider the case, when
there are no poles and zeros. The corresponding example of the lattice model is
given by Wilson fermions and has been considered in \cite{VZ2012}. In the next
section the consideration will be extended to a more general case.

\subsection{Invariant $\tilde{\cal N}_5$}

Let us consider the Euclidean Green's function on the 4D lattice ${\cal G}$ as
the inverse Hamiltonian in 4D momentum space and introduce the 5D Green's
function:
\begin{equation}
\label{Green5}
 G^{-1}(p_5,p_4,{\bf p})= p_5 \gamma^5 + {\cal G}^{-1}(p_4,{\bf p}) = (i p_5  + {\cal Q}^{-1}(p_4,{\bf
 p}))(-i \gamma^5)
\,.
\end{equation}
Then one can introduce the topological invariant as  the 5-form  (see also
\cite{Kaplan2011,Kaplan1992,Golterman1993,Volovik2003,SilaevVolovik2010,ZhongWang2010}):
\begin{definition}
\begin{equation}
\label{N_5} {\cal N}_5 = \frac{1}{2 \pi^3 5! i} {\rm Tr}\,  \int G d G^{-1}
\wedge G d G^{-1}\wedge G d G^{-1} \wedge G d G^{-1}\wedge G d G^{-1} \,,
\end{equation}
where the integration is over the  Brillouin zone in 4D momentum space
$(p_4,{\bf p})$ and over the whole $p_5$ axis.
\end{definition}

The properties of this invariant are summarized in the following lemma (see
also \cite{Z2011}, where the similar $3D$ construction is considered):
\begin{lemma}
Eq. (\ref{N_5}) defines the topological invariant for the gapped $4D$ system
with momentum space $\cal M$ if the following equation holds:
\begin{equation}
\int_{\partial [{\cal M}\otimes R]}  \, {\bf tr} \left( [\delta {\cal G}^{-1}]
{ G} \,
  d  { G}^{-1}\wedge
  d  { G}\wedge d  { G}^{-1}\wedge
  d  { G}\right)=0,\quad  p_5^2 \rightarrow \infty \label{dN5}
\end{equation}
 This requirement is satisfied, in particular, for the system with compact closed
$\cal M$.

For the system with the Green function of the form (\ref{G_}) expression for
the topological invariant is reduced to
\begin{equation}
\label{N_51} \tilde{\cal N}_5 = \frac{3}{4 \pi^2 4!} \epsilon_{abcde}\, \int
\hat{g}^a\, d \hat{g}^b \wedge d \hat{g}^c \wedge d \hat{g}^d \wedge d
\hat{g}^e,\quad \hat{g}^a = \frac{g^a}{\sqrt{g^cg^c}}
\end{equation}
\end{lemma}

(This lemma corresponds to the Theorem from Sect. 4.3 of \cite{VZ2012}.)  
The important particular cases, when Eq. (\ref{N_5}) defines the topological invariant are also listed in \cite{VZ2012}. 
Requirement (\ref{dN5}) is satisfied, in particular, for the system with compact
closed momentum space and for the system with the green function that is massless in
ultraviolet. In particular, the model with the fermions coupled to the asymptotic free gauge fields satisfy  (\ref{dN5}) .

\subsection{Topological invariant $\tilde{\cal N}_4$}

In addition to the invariant $\tilde{\cal N}_5$ let us also consider a
different construction that coincides with  $\tilde{\cal N}_5$ for the case of
free fermions
\begin{definition}
\begin{eqnarray}
\tilde{\cal N}_4 &=& \frac{1}{48 \pi^2} {\rm Tr}\, \gamma^5 \int_{\cal M}
d{\cal G}^{-1}\wedge d {\cal G} \wedge d {\cal G}^{-1} \wedge d {\cal G}
\label{N40}
\end{eqnarray}
Here the integration is over the whole $4D$ space $\cal M$.
\end{definition}

 The expression in
this integral is the full derivative. That's why the given invariant can be
reduced to the integral over the $3D$ hypersurface $\partial {\cal M}$:
\begin{eqnarray}
\tilde{\cal N}_4 & = & \frac{1}{48 \pi^2} {\rm Tr}\, \gamma^5 \int_{\partial
{\cal M}} {\cal G}^{-1} d {\cal G} \wedge d {\cal G}^{-1} \wedge d {\cal
G}\label{N40_}
\end{eqnarray}

 The
last equation is identical to that of for the invariant ${\cal N}_3$ for
massless fermions. Therefore, Eq. (\ref{N40}) defines the topological invariant
if the Green function anticommutes with $\gamma^5$ on the boundary of momentum
space. In particular, for the noninteracting fermions
 $\tilde{\cal N}_4 = {\rm Sp}\, {\bf 1}$ (the number of Dirac
fermions).

The following lemma allows to calculate invariant $\tilde{\cal N}_4$ in general
case:

\begin{lemma}
\label{calcN4} For the Green function of the form (\ref{G_}) with
$\frac{m[p]}{\sqrt{g_a g_a+m^2}} = 0\, (a = 1,2,3,4)$ on $\partial{\cal M}$ we
have:
\begin{equation}
\label{N_41} \tilde{\cal N}_4 = \frac{1}{2 \pi^2 3!} \epsilon_{abcd}\,
\int_{\partial{\cal M}} \hat{g}^a\, d \hat{g}^b \wedge d \hat{g}^c \wedge d
\hat{g}^d,\quad \hat{g}^a = \frac{g^a}{\sqrt{g^cg^c}}
\end{equation}
\end{lemma}
(This Lemma follows when one substitutes Eq. (\ref{G_}) to Eq. (\ref{N40_}).)

\subsection{Pseudo - poles of the Green function}
\label{N5calc}

Let us introduce the following parametrization
\begin{equation}
\hat{g}_5 = {\rm cos} 2 \alpha, \quad \hat{g}_a = k_a {\rm sin} 2 \alpha
\end{equation}

Vector $k$ may be undefined at the points of momentum space ${\cal M}$, where
$\hat{g}^a = 0, a = 1,2,3,4$. In nondegenerate case this occurs on points $y_i,
i = 1, ...$. Further we call these points the pseudo - poles of the Green
function.
\begin{definition}
The point $y_i$ in momentum space, where $g^a = 0,\, a = 1,2,3,4$ and,
therefore, ${\cal G}^{-1}[p] = m[p]$, is called pseudo - pole of the Green
function.
\end{definition}

Actually, in the majority of cases at these points massive fermion excitations
appear. This is because in these cases (free continuum fermions, lattice Wilson
fermions, overlap fermions, etc) for infinitely small $m[y_i]$ needed to
approach continuum limit, the Green function behaves as
\begin{equation}
{\cal G} \sim \frac{1}{\lambda \,\sum_a (-1)^{n_a} q_a \gamma^a - i m},
\end{equation}
where $q_a = p_a - y_i$,   $\lambda$ is a real constant, $n_a$ are integer
constants. Therefore, in Minkowsky space the usual dispersion relation is
recovered: $E = \sqrt{q^2 + m^2/\lambda^2}$. We formulate this as the following
lemma:
\begin{lemma}
If $g^a \sim \lambda \sum_a (-1)^{n_a} (p^a - y^a)$ while $m\ne 0$ in the small
vicinity of $y \in {\cal M}$, then at this point massive fermion excitation
appears.
\end{lemma}

Let us denote a small vicinity of the pseudo - pole $y_i$ by $\Omega(y_i)$.
According to \cite{VZ2012} we have
\begin{eqnarray}
\tilde{\cal N}_5 &=&  \frac{1}{ \pi^2 4!} \epsilon_{abcd}\, \int_{\sum_{i =
0,1,...}\partial \Omega(y_i)-\partial {\cal M}} (3 \hat{g}_5 - \hat{g}_5^3 )
k^a\, d k^b \wedge d k^c \wedge d k^d \label{N5p}
\end{eqnarray}
 Let us define the $3D$ analogue of the residue.
\begin{definition}
We denote by ${\bf Res}(p)$ the degree of mapping $\{\hat{g}^a: S^3 \rightarrow
S^3\}$:
\begin{eqnarray}
{\bf Res}(p) &=&  \frac{1}{ 2 \pi^2 3!} \epsilon_{abcd}\, \int_{\partial
\Omega(p)} \hat{g}^a\, d \hat{g}^b \wedge d \hat{g}^c \wedge d
\hat{g}^d,\nonumber\\ && p \in \Omega, \quad |\Omega| \rightarrow 0
 \label{CI}
\end{eqnarray}
\end{definition}

Let us also denote
\begin{equation}
{\bf s}(p) = {\rm sign}\, m[p]
\end{equation}

The lemma follows:
\begin{lemma}
\label{calcN5}  Consider the system with the Green function of the form
(\ref{G_}), with $\frac{m[p]}{\sqrt{g_a g_a+m^2}} = 0\, (a = 1,2,3,4)$ on the
boundary of momentum space. Let us denote by $y_i$ the points in momentum
space, where $g^a = 0, \, a = 1,2,3,4$. Then  the topological invariant
$\tilde{\cal N}_5$ is given by
\begin{eqnarray}
\tilde{\cal N}_5 &=&  \sum_{i = 0,1,...} {\bf s}(y_i) {\bf Res}(y_i)
\label{N5f}
\end{eqnarray}
\end{lemma}
(This is Eq. (31) of \cite{VZ2012}.)
For the details of the proofs of the lemmas presented in this section see \cite{VZ2012}.

\section{The topological invariants in general case and index theorem} \label{SectIndTheor}

\subsection{Topological invariants for the case when $\cal G$ has zeros and poles}

 In this section we generalize the definition of the topological invariants to the case, when the Green function may have
 zeros or poles. Construction of invariants $\tilde{\cal N}_4$ and $\tilde{\cal N}_5$ in
 this case requires some care. The correct definition implies that
 first we consider momentum space without some vicinities $\Omega (z_i), \Omega (p_i)$ of the points $z_i$, where
 $\cal G$ has zeros and points $p_i$, where there are poles of the Green function. 
Then all statements of the previous section (proved in \cite{VZ2012}) are valid if we consider the Green function in Momentum space without  $\Omega (z_i), \Omega (p_i)$. That's why, say, in Eq. (\ref{N5p}) and Eq. (\ref{N_41}) of  the previous section we must add $-\sum_i\partial \Omega (z_i) -\sum_j\partial \Omega (p_j)$ to $\partial {\cal M}$. 

 Next, the limit is considered when the sizes of these
 vicinities tend to zero $|\Omega(p_i)|,|\Omega(z_i)| \rightarrow 0$. In order for such a limit to exist the model must
 obey some requirements.

\begin{remark}
In this paper we consider only the cases, when exceptional points of the Green
function are indeed point - like. We do not consider the situation when
exceptional lines or surfaces of the Green function are present. (The case of
the exceptional surface may correspond, in particular, to the Fermi surface.)
\end{remark}

 For example, if the Green function has the form Eq. (\ref{G_}) and $\hat{g}_5 = \frac{m}{\sqrt{g_a g_a+m^2}} = 0\, (a =
 1,2,3,4)$ at $z_i,p_i$, then the boundaries $\partial \Omega(p_i), \partial \Omega(z_i)$ do not
 contribute to the sum in (\ref{N5p}) at $|\Omega(p_i)|, |\Omega(z_i)| \rightarrow 0$. This
 means that the mentioned above limit exists for $\tilde{\cal N}_5$. At the same time under the same
 conditions $\tilde{\cal N}_4$ is the topological invariant at $|\Omega(p_i)|,|\Omega(z_i)| \rightarrow 0$ and
 the points $p_i, z_i$ contribute the sum in Eq. (\ref{N_41}).

In order to make formulas more simple let us also introduce the $3D$ residue at
"infinity":
\begin{eqnarray}
{\bf Res}(\infty) &=&  -\frac{1}{ 2 \pi^2 3!} \epsilon_{abcd}\, \int_{\partial
{\cal M}} \hat{g}^a\, d \hat{g}^b \wedge d \hat{g}^c \wedge d \hat{g}^d
\end{eqnarray}

 Taking into account Lemma \ref{calcN5} and Lemma \ref{calcN4}, we come to the following
\begin{theorem}
\label{N54theorem} Suppose the Green function has the form (\ref{G_}) and at
its poles and zeros as well as on the boundary of momentum space
$\frac{m[p]}{\sqrt{g_a g_a+m^2}} = 0\, (a = 1,2,3,4)$. We denote by $y_i$ the
points, where ${\cal G}^{-1}(y_i) = m(y_i)$ (pseudo - poles of $\cal G$), by
$z_i$ the points, where ${\cal G}(z_i) = 0$, by $p_i$ the points, where ${\cal
G}^{-1}(z_i) = 0$. Then $\tilde{\cal N}_4$ and $\tilde{\cal N}_5$ are well
defined topological invariants. (This means that $\tilde{\cal N}_4$ and $\tilde{\cal N}_5$ are not changed 
under the smooth deformation of $\cal G$ that keeps the listed above conditions.) As a result we have 
\begin{eqnarray}
\tilde{\cal N}_5 &=&  \sum_{i = 0,1,...} {\bf s}(y_i) {\bf Res}(y_i)
\label{N5f__}
\end{eqnarray}
and
\begin{eqnarray}
\tilde{\cal N}_4 &=& - \sum_{i = 0,1,...} {\bf Res}(z_i) -  \sum_{i = 0,1,...}
{\bf Res}(p_i) -  {\bf Res}(\infty)\label{N4f_}
\end{eqnarray}
\end{theorem}

\subsection{Generalized unparticles and Index theorem}

In addition to zeros and poles in general case momentum space may contain non -
analytical exceptional points $q_i$, where $\cal G$ is not defined but, say,
${\cal G}^{-1}$ may differ from zero.
\begin{definition}
The point $q_i$ in momentum space represents generalized unparticle if in its
small vicinity both $\cal G$ and ${\cal G}^{-1}$ are not analytical as
functions of momenta.
\end{definition}

As usual, poles $p_i$ of $\cal G$ (such points that ${\cal G}^{-1}$ is zero at
$p_i$ but remains analytical in its vicinity) represent massless particles.

\begin{remark} If momentum space contains generalized unparticles, then both $\tilde{\cal N}_5$ and
$\tilde{\cal N}_4$ are not well - defined.
\end{remark}

The pattern of the transition from the state at $\beta> \beta_c$ to the state
at $\beta < \beta_c$ can be described in terms of the flow of exceptional
points of $g^a$. Namely, in general there are zeros $y_i$ of $g^a$, where
${\cal G}^{-1} = m$ (we call them pseudo - poles, some of these points become
poles $p_i = y_i$ of ${\cal G}$ if, in addition, $m[y_i] = 0$). Also there are
zeros $z_i$ of ${\cal G}$, where $g^a \rightarrow \infty$. These points cannot
simply disappear when the system is changed smoothly with no phase transition encountered.
 They may annihilate each other if this is allowed by the
momentum space topology. Namely, two zeros $z_i$, $z_j$ may annihilate if ${\bf
Res}(z_i) + {\bf Res}(z_j) = 0$ because in this case they do not contribute the
sum in Eq. (\ref{N4f_}). For the same reason two poles $p_i$, $p_j$ may
annihilate if ${\bf Res}(p_i) + {\bf Res}(p_j) = 0$.  Two pseudo - poles $y_i,
y_j$ may annihilate each other if ${\bf s}(y_i){\bf Res}(y_i) + {\bf
s}(y_i){\bf Res}(y_j) = 0$ because in this case they do not contribute the sum
in Eq. (\ref{N5f__}).

Now we are ready to formulate the generalized index theorem:
\begin{theorem}
\label{indextheorem} Suppose that the $4D$ system with the Green function of
the form (\ref{G_}) depends on parameter $\beta$ and there is a phase
transition at $\beta_c$ with changing of $\tilde{\cal N}_4$ and $\tilde{\cal
N}_5$. At $\beta \ne \beta_c$ the system does not contain generalized
unparticles and massless excitations. $\cal G$ as a function of $\beta$ is
smooth everywhere except for the points, where it is not defined (even at the phase transition). Green
function may have zeros $z_i$, and $m[z_i] \ne \infty$.    Momentum space $\cal
M$ of the $4D$ model is supposed to be either compact and closed or open. In
the latter case we need that $\cal G$ does not depend on $\beta$ on $\partial
{\cal M}$. Then at $\beta = \beta_c$ the number of topologically protected
generalized unparticles is
\begin{equation}
n^u_f  = \Delta \tilde{\cal N}_4 \label{nuz}
\end{equation}
The number of topologically protected massless excitations at $\beta_c$ is
\begin{equation}
n^0_f = \frac{1}{2}\Delta \tilde{\cal N}_5 - \frac{1}{2} \{\sum_{i: z_j
\rightarrow y_i}  {\bf s}(y_i) \, {\bf Res}(y_i)|_{\beta > \beta_c} -\sum_{i:
y_i \rightarrow z_j} {\bf s}(y_j) \, {\bf Res}(y_j)|_{\beta < \beta_c}
\}\label{nfz}
\end{equation}
Here the first sum is over the pseudo - poles that are transformed to zeros at
$\beta_c$ while the second sum is over the zeros that become pseudo - poles.

\end{theorem}


{\bf Proof} (See also the discussion before the statement of the theorem.)

We made here an important supposition that  $\cal G$ as a function of $\beta$ is smooth everywhere except for the points, where it is not defined, even at the phase transition. Therefore, the behavior of the Green function may be predicted even at the position of the phase transition.  Namely, at the phase transitions the exceptional points of the Green function also cannot disappear or arise if they are topologically protected, i.e. if ${\bf Res}$ at the corresponding points is nonzero.  (Otherwise the Green function would experience a non - smooth change over all momenum space.)  
Thus, the exceptional points of different types may be
transformed to each other at the phase transition point $\beta = \beta_c$. It will be shown below that at
the position of the transformation of the zero to the pseudopole the non - analyticity of $\cal G$ takes
place, i.e. the generalized unparticle appears.

Let us consider the flow of the zeros $z_i$ of $\cal G$ in $\beta$. In fact,
the zeros of the Green function with opposite signs of the corresponding terms
in (\ref{N4f_}) may annihilate while the zeros with the same signs may not. (See the discussion before the statement of the theorem.) If
$\Delta{\cal N}_4 \ne 0$ then there is the corresponding number of exceptional
points for the Green function at $\beta_c$. These points mark the transformation of
the zeros into the pseudo - poles. $\cal G$ is defined at both these kinds of
exceptional points. Therefore, $\cal G$ remains analytical function of $\beta$
at the transition point in the vicinities of the given points. Consider, for
example, the case, when the zero $z$ becomes pseudo - pole at $\beta = \beta_c
-\epsilon \rightarrow \beta_c + \epsilon$. Suppose that it is possible to
expand $\cal G$ as a series in powers of $x = p - z$ in small vicinity of
$\beta$. Then at $\beta > \beta_c $:
\begin{eqnarray}
{\cal G}(p) & \sim & \gamma^a [f_{b}^a x^b + \frac{1}{2!} f^a_{bc} x^b x^c +
...] + [r + r_{b}x^b + \frac{1}{2!} r_{bc} x^b x^c + ...]
\end{eqnarray}
with some coefficients $f,r$ that depend on $\beta$. At the same time at $\beta
< \beta_c $
\begin{eqnarray}
{\cal G}(p) & \sim & \gamma^a [f_{b}^a x^b + \frac{1}{2!} f^a_{bc}x^b x^c +
...] + [ \frac{1}{2!} r_{bc}x^b x^c + ...]
\end{eqnarray}
Here $r$ and $r_b$ vanish because the Green function has zero and at this zero
$\hat{g}_5 = \frac{m}{g^2 + m^2}= 0$. This becomes obvious when $g^a, m$ are
written in terms of $f,r$:
\begin{eqnarray}
g^a(p) & \sim & \frac{f_{b}^a x^b}{f^a_{b}f^a_{c} x^b x^c }, \quad  m(p) \sim
\frac{\frac{1}{2} r_{bc} x^b x^c}{f^a_{b}f^a_{,c} x^b x^c } \quad
(\beta < \beta_c)\nonumber\\
g^a(p) & \sim & \frac{1}{r^2} f_{b}^a x^b, \quad m(p) \sim \frac{1}{r} \quad
(\beta
> \beta_c)
\end{eqnarray}
By condition all functions $f, r$ are analytic in $\beta$. We come to the
contradiction: functions $r,r_b$ cannot be analytic at $\beta_c$ because at
$\beta < \beta_c$ functions $r(\beta)$ and $r_b(\beta)$ vanish while at $\beta
> \beta_c$ do not. This shows that at $\beta_c$ both series are undefined.
That's why we come to the conclusion that the exceptional point is non -
analytical at $\beta_c$. The number $\Delta \tilde{\cal N}_4$ measures the
change in the number of zeros across the transition. The zero disappears when
it becomes the pseudo - pole. We come, therefore to Eq. (\ref{nuz}).

In order to prove the second statement of the theorem  let us notice that the
topologically protected poles cannot appear instantaneously. Namely, the pseudo
- poles of $\cal G$ existing at $\beta \ne \beta_c$ may become the poles of
$\cal G$ at $\beta = \beta_c$. This happens if $m(y_i)$ changes sign at
$\beta_c$.  Also in general case the pseudo - pole may be transformed to zero
of the Green function through the formation of the generalized unparticle in
the intermediate state. That's why the total algebraic number of topologically
protected massless fermions in the intermediate state  is given by
\begin{equation}
n_f^0 = \frac{1}{2} \sum_{i, y_i \not \rightarrow z_j} \Delta {\bf s}(y_i) \,
{\bf Res}(y_i)
\end{equation}
Here the sum is only over those pseudo - poles that are not transformed to
zeros across the transition.  We then come to Eq. (\ref{nfz}) in an obvious
way.

\begin{remark}
There are two important particular cases:


1. If zeros are not transformed to pseudo - poles and vice versa, then
\begin{equation}
n^0_f = \frac{1}{2}\Delta \tilde{\cal N}_5 \label{nfz0}
\end{equation}

2. If ${\rm sign} (m)$ remains constant on $\cal M$ and as a function of
$\beta$, then $n^0_f = 0$ and
\begin{equation}
n^0_f = \frac{1}{2}\Delta \{ \tilde{\cal N}_5 - {\bf s} \tilde{\cal N}_4\} =
0\label{nfz1}
\end{equation}
From here we obtain $\Delta \tilde{\cal N}_5 = {\rm sign}( m) \, \Delta
\tilde{\cal N}_4$ if ${\rm sign}(m) = const$. In the other words,
\begin{eqnarray}
\Delta\{ \sum_{i = 0,1,...} {\bf Res}(z_i) +  \sum_{i = 0,1,...} {\bf
Res}(y_i)\} & = & 0 \label{C0}
\end{eqnarray}


\end{remark}

\begin{remark}
The total observed number of generalized unparticles and massless fermions may
be larger than $n_f^u$ and $n_f^0$. In this case some of these excitations may
annihilate each other without a phase transition.
\end{remark}

In a similar way the following theorem may be proved

\begin{theorem}
\label{indextheorem2} Suppose that the $4D$ system with the Green function of
the form (\ref{G_}) depends on parameter $\beta$ and there is a phase
transition at $\beta_c$ with changing of $\tilde{\cal N}_4$ and $\tilde{\cal
N}_5$. At $\beta \ne \beta_c$ all excitations are massless. $\cal G$ as a
function of $\beta$ is smooth everywhere except for the points, where it is not
defined. Green function may have zeros $z_i$, and $m[z_i] \ne \infty$. Momentum
space $\cal M$ of the $4D$ model is supposed to be either compact and closed or
open. In the latter case we need that $\cal G$ does not depend on $\beta$ on
$\partial {\cal M}$. Then at $\beta = \beta_c$ the number of topologically
protected generalized unparticles is
\begin{equation}
n^u_f  = \Delta \tilde{\cal N}_4 \label{nuz}
\end{equation}
\end{theorem}

\section{Overlap fermions}
\label{lattice}

\subsection{Wilson kernel}

In this section we consider the case of the single flavor of lattice Dirac
fermions. In lattice regularization the periodic boundary conditions are used
in space direction and antiperiodic boundary conditions are used in the
imaginary time direction. The momenta to be considered, therefore, also belong
to a lattice:
\begin{equation}
p_a = \frac{2\pi K_a}{N_a}\,  \quad p_4 = \frac{2\pi K_4+\pi}{N_t}, \quad
K_a,K_4 \in Z\quad a = 1,2,3
\end{equation}
Here $N_a, N_t$ are the lattice sizes in $x$, $y$, $z$, and imaginary time
directions, correspondingly. However, for the lattice of infinite volume we
recover continuous values of momenta that belong to the $4D$ torus.

For the free Wilson fermions the Green function has the form \cite{Montvay}:

\begin{eqnarray}
{\cal G}& = & \Bigl( \sum_a \gamma_a {\rm sin}\, p_a - i (m + \sum_a (1 - {\rm
cos}\, p_a)) \Bigr)^{-1}\nonumber\\ & = & \frac{ \sum_a \gamma_a {\rm sin}\,
p_a + i (m + \sum_a (1 - {\rm cos}\, p_a)) }{\sum_a {\rm sin}^2\, p_a + (m +
\sum_a (1 - {\rm cos}\, p_a))^2}, \quad a = 1,2,3,4
\end{eqnarray}

\begin{table}
\begin{center}
\begin{small}
\begin{tabular}{|c|c|c|c|c|c|c|c|c|c|c|c|c|c|c|c|c|}
\hline
$m$ & $\tilde{\cal N}_4$ & $\tilde{\cal N}_5$   \\
\hline
$m>0 $ & $0$ & $0$ \\
\hline
$-2 < m < 0$ & $0$ & $-2$  \\
\hline
$-4 < m < -2 $ & $0$ & $6$ \\
\hline
$-6 < m < -4$ & $0$ & $-6$  \\
\hline
$-8 < m < -6 $ & $0$ & $2$  \\
\hline
$ m < -8 $ & $0$  & $0$ \\
\hline
\end{tabular}
\end{small}
\end{center}
\caption{The values of topological invariants $\tilde{\cal N}_4$ and
$\tilde{\cal N}_5$ for free Wilson fermions. } \label{table2}
\end{table}

In Table \ref{table2} the values of $\tilde{\cal N}_5$ for the Wilson fermions
calculated in \cite{VZ2012} are presented. The index theorem states that the
total number $n_F$ of gapless fermions emerging at the critical values of mass
$m$ is determined by the jump in $\tilde{\cal N}_5$:
\begin{equation}
\label{IndexTheorem} n_F=\frac{1}{2}\Delta\tilde{\cal N}_5
\end{equation}
We also have an explicit formula for the topological invariant
\cite{VZ2012,Kaplan1992,Golterman1993}
\begin{equation}
\tilde{\cal N}_5= \sum_{k=0}^{4} (-1)^k C_4^k \frac{m+2k}{|m+2k|},
\end{equation}
where the sum is over the fermion doublers in $4D$, $m + 2k$ is the mass of the
$k$-th doubler while $C_4^k$ is its degeneracy. It is worth mentioning that
$\tilde{\cal N}_4 = 0$ for Wilson fermions.

When the interaction of Wilson fermions with the lattice gauge field ${\cal U}
= e^{i {\cal A}}$ defined on links is turned on, we have (in coordinate space):
\begin{eqnarray}
&&{\cal G}(x,y)  =  \frac{i}{Z} \int  D{\cal U}\, {\rm exp} \Bigl( - S_G[{\cal
U}] \Bigr) \, {\rm Det}  ({\cal D}[{\cal U},m]) {\cal D}_{x,y}^{-1}[{\cal U},m]
\label{Gr}
\end{eqnarray}
where $S_G$ is the gauge field action while
\begin{equation}
{\cal D}_{x,y}[{\cal U},m]  =  - \frac{1}{2}\sum_i [(1 +
\gamma^i)\delta_{x+{\bf e}_i, y} {\cal U}_{x+{\bf e}_i, y}  +  (1 -
\gamma^i)\delta_{x-{\bf e}_i, y} {\cal U}_{x-{\bf e}_i, y}] +  (m + 4)
\delta_{xy}
\end{equation}

Here ${\bf e}_i$ is the unity vector in the $i$ - th direction. Again, the
Green function in momentum space is expected to  have the form \cite{Shrock} of
Eq. (\ref{G_}).

\subsection{Basics of overlap fermions}

\begin{table}
\begin{center}
\begin{small}
\begin{tabular}{|c|c|c|c|c|c|c|c|c|c|c|c|c|c|c|c|c|}
\hline
$m_0 < 0 $ & -  & -  & -  & -  &  - \\
\hline
$2 > m_0 > 0 $ & $1\otimes m$  & -  & -  & -  &  - \\
\hline
$4 > m_0 > 2$ & $1\otimes m$  & $4\otimes [m(1 - \frac{2}{m_0} )]$ & -  & -  &  - \\
\hline
$6 > m_0 > 4$ & $1\otimes m$  & $4\otimes [m(1 - \frac{2}{m_0} )]$ & $6\otimes [m(1 - \frac{4}{m_0} )]$  & -  &  - \\
\hline
$8 > m_0 > 6$ & $1 \otimes m$  & $4 \otimes [m(1 - \frac{2}{m_0} )]$ & $6\otimes [m(1 - \frac{4}{m_0} )]$  & $4\otimes [m(1 - \frac{6}{m_0} )]$  &  - \\
\hline
$m_0 > 8$ & $1\otimes m$  & $4\otimes [m(1 - \frac{2}{m_0} )]$ & $6\otimes [m(1 - \frac{4}{m_0} )]$  & $4\otimes [m(1 - \frac{6}{m_0} )]$  &  $1\otimes [m(1 - \frac{8}{m_0} )]$ \\
\hline
\end{tabular}
\end{small}
\end{center} \caption{The spectrum of the system with free overlap fermions. In the first column the values of $m_0$ are specified. In
the other columns masses of the doublers are listed.  Expression $x \otimes {v}$ means
$x$ states with the masses equal to $v$. } \label{table5}
\end{table}

Below we consider briefly the properties of overlap fermions. This type of the
regularization is now widely used in the numerical simulations because it
provides a lattice version of the continuum chiral symmetry that is lost in
simpler forms of the regularization like the Wilson fermions considered above.
In this regularization the propagator has the form:
\begin{eqnarray}
&&{\cal G}(x,y)  =  \frac{1}{Z} \int D D{\cal U}\, {\rm exp} \Bigl( -
\tilde{S}_G[{\cal U}] \Bigr) \, \{-i{\bf D}[{\cal U}] -  i m\}_{x,y}^{-1}
\label{Gr}
\end{eqnarray}
Here the effective action $\tilde{S}_G$ includes also the fermion determinant,
the overlap operator is defined as
\begin{equation}
{\bf D} = \frac{2m_0}{{\cal O}^{-1}-1}
\end{equation}
with
\begin{eqnarray}
{\cal O}_{x,y}[{\cal U}]  & = &  \frac{1}{2}\Bigl( 1 + \frac{{\cal D}[{\cal
U},-m_0]}{\sqrt{{\cal D}^+[{\cal U}, - m_0] {\cal D}[{\cal U},-m_0]}}\Bigr)
\end{eqnarray}
Here $m_0$ and  $m$ are bare mass parameters \cite{Overlap}. The parameter $m$
represents bare physical mass.

In spite of a rather complicated form of the expression for the overlap
operator it is commonly believed that the Green function in momentum space has
the same form (\ref{G_}) as the Green function for Wilson fermions
\cite{Overlap}. In particular, for the free overlap fermions the Green function
has the form:
\begin{equation}
{\cal G}(p) = \frac{1}{g^i[p] \gamma^i - im}
\end{equation}
with (see Appendix in \cite{Overlap}):
\begin{eqnarray}
g^i[p] & = & 2 m_0 \, {\rm sin}\, p^i \, \frac{A(p) + \sqrt{A(p)^2+\sum_i{\rm
sin}^2 \, p^i}}{\sum_i {\rm sin}^2 \,
p^i}, \nonumber\\
A(p) &=& -m_0 + \sum_i [1 -  {\rm cos}\, p^i]
\end{eqnarray}

In Table \ref{table5} we represent the spectrum of the model for different
values of $m_0$. For $0 < m_0 < 2$ one obtains that $\hat{g}^a$ have the only
zero at $p = 0$. However, there are poles of $\hat{g}^a$ at $p_{n_i} = (\pi
n_1, \pi n_2, \pi n_3, \pi n_4), \quad n_i = 0,1, \, \sum n^2 \ne 0$. The
values of $\hat{g}_5$ at these points vanish.

 In general case some of the poles of $\hat{g}^a$ may become zeros depending on the value of $m_0$.
Near to these poles(zeros) we have:
\begin{equation}
g^i[p_{n_i}+\delta p]  =  2 m_0 \, (-1)^{n_i} \, \frac{\delta p^i}{{\sum_i
[\delta p^i]^2}} \, \Bigl( \frac{1}{2  |A(p_{n_i})| } \sum_i [\delta p^i]^2
  + (A(p_{n_i}) + | A(p_{n_i})|)    \Bigr)
\end{equation}

For positive $A(p_{n_i})$ we obtain $g^i[p_{n_i}+\delta p] \sim 4  m_0
\,A(p_{n_i}) (-1)^{n_i} \, \frac{\delta p^i}{|\delta p|^2} $, where $|\delta
p|^2 = \sum_i [\delta p^i]^2 $ and
\begin{equation}
{\cal G}(p_{n_i}+\delta p) \sim \frac{\frac{1}{4  m_0 \,A(p_{n_i})} |\delta
p|^2}{\sum_i (-1)^{n_i} \, \delta p^i \,  \gamma^i - i \frac{m}{4 m_0
\,A(p_{n_i})} |\delta p|^2}\sim \frac{1}{4  m_0 \,A(p_{n_i})} \sum_i (-1)^{n_i}
\, \delta p^i \,  \gamma^i \label{A+}
\end{equation}
We have zeros of the Green function at these points.

For negative  $A(p_{n_i})$ we obtain $g^i[p_{n_i}+\delta p] \sim
\frac{m_0}{|A(p_{n_i})|}  (-1)^{n_i} \, \delta p^i$ and
\begin{equation}
{\cal G}(p_{n_i}+\delta p) \sim \frac{\frac{|A(p_{n_i})|}{ m_0 }}{\sum_i
(-1)^{n_i} \, \delta p^i \,  \gamma^i - i m \frac{|A(p_{n_i})|}{ m_0
}}\label{A-}
\end{equation}
  The values  $A(p_{n_i}) = -m_0 + 2 \sum_i n_i$ at these points are related to
the values of the masses of the doublers: ${\bf m}_{n_i} = m ( 1 -
\frac{2}{m_0} \sum_i n_i ) $.

The special situation appears if $A(p_{n_i}) = 0$ (this occurs for the
intermediate values $m_0 = 2,4,6,8$):
\begin{equation}
{\cal G}(p_{n_i}+\delta p) \sim \frac{\frac{1}{ 2m_0 }}{\sum_i (-1)^{n_i} \,
\frac{\delta p^i}{|\delta p|} \,  \gamma^i - i  \frac{m}{2 m_0 }},\quad |\delta
p| = \sqrt{\sum_i [\delta p^i]^2}\label{A0}
\end{equation}
In this case the Green function is not defined at  $p_{n_i}$ and the
unparticles appear with the propagator equal (up to the normalization constant)
to that of presented in \cite{fermion_unparticle} (follows from Eq. (8) of
\cite{fermion_unparticle} with $\alpha = \beta = 0, \zeta \ne 0, d_U = 2$). At
$m = 0$ we arrive at the propagator given in Eq. (10) of
\cite{fermion_unparticle} with $\alpha = 0, d_U =2$.

\subsection{$\tilde{\cal N}_4$ and $\tilde{\cal N}_5$ for overlap fermions}

It is worth mentioning that for the overlap fermions unlike the Wilson fermions
there are zeros of the Green function at some points in momentum space (see Eq.
(\ref{A+})). Moreover, at the intermediate values of $m_0$ the Green function
is undefined at some points (see Eq. (\ref{A0} ) ). As a result, we need to
consider momentum space without small vicinities of both mentioned types of the
points in order to calculate $\tilde{\cal N}_5$ and $\tilde{\cal N}_4$.
Momentum space, therefore, becomes open. At $m_0 \ne 0,2,4,6,8$ we have at the
points, where $\cal G$ has zeros $\hat{g}_5 = 0$ due to Eq. (\ref{A+}).
Therefore, the conditions of Theorem \ref{N54theorem} are satisfied. For the
intermediate values of $m_0$ this theorem cannot be applied.

For $0 < m_0 < 2$ only the physical value $p = 0$ contributes the sum in Eq.
(\ref{N5f}). The sign of $\hat{g}_5$ is positive. As a result $\tilde{\cal N}_5
= {\rm sign}\, m $ for $0 < m_0 < 2$.

 For $2 < m_0 < 4$ there exist $5$ modes with negative $A$ at
$p = (\pi, 0, 0 ,0); p = (0,\pi, 0 ,0); p = (0, 0, \pi, 0); p = (0 , 0, 0
,\pi); p = (0, 0, 0 ,0)$ and we have $\tilde{\cal N}_5 = (1 - 4){\rm sign} \, m
= -3 \, {\rm sign}\, m$.

In the general case we have:
\begin{equation}
\tilde{\cal N}_5= \frac{1}{2}{\rm sign}\, m\, \sum_{k=0}^{4} (-1)^k C_4^k
\Bigl( \frac{m_0-2k}{|m_0-2k|} - 1 \Bigr) = -\frac{1}{2}{\rm sign}\, m\,
\sum_{k=0}^{4} (-1)^k C_4^k \frac{-m_0+2k}{|-m_0+2k|}
\end{equation}

In order to calculate $\tilde{\cal N}_4$ we also use Theorem \ref{N54theorem}.
In a similar way to the above calculation of $\tilde{\cal N}_5$ we have:
\begin{equation}
\tilde{\cal N}_4 = - \frac{1}{2}\sum_{k=0}^{4} (-1)^k C_4^k
\frac{-m_0+2k}{|-m_0+2k|} = \tilde{\cal N}_5\,{\rm sign}\, m
\end{equation}

The values of $\tilde{\cal N}_4$ and  $\tilde{\cal N}_5$ for overlap fermions
versus parameter $m_0$ are represented in Table \ref{table4}. Let us remind
that unlike the free Wilson fermions at intermediate values of the mass
parameter $m_0 = 0, 2, 4, 6 ,8$ there exist exceptional points in momentum
space such that the Green function is undefined at these points. Due to this
the invariants $\tilde{\cal N}_4$ and $\tilde{\cal N}_5$ are not well defined
in the intermediate states.

When the interaction with the gauge fields is turned on, it is necessary to
check that $\tilde{\cal N}_4$ and $\tilde{\cal N}_5$ remain topological
invariants. First, $\tilde{\cal N}_4$ remains topological invariant if $\{
\gamma^5, \delta {\cal G}\} = 0$ on $\partial {\cal M}$. In our case $\partial
{\cal M}$ encloses zeros of $\cal G$. That's why the question is: does
$\hat{g}_5$ vanish at the points, where ${\cal G}=0$, or not. Near to these
points without interactions we have the expression for $\cal G$ given by Eq.
(\ref{A+}). When the interactions are turned on, this expression is changed.
However, under an assumption that it still remains analytic in $\delta p$, we
obtain $g_5[p] \sim \delta p^2$ while $g_a \sim \frac{\delta p_a}{p^2}$.
Therefore, at $\delta p \rightarrow 0$ we recover $\hat{g}_5 \rightarrow 0$.
This kind of behavior is expected until the phase transition is encountered.
So, we come to the conclusion that $\tilde{\cal N}_4$ remains the topological
invariant when the interactions are turned on.

In order for the functional $\tilde{\cal N}_5$ to remain the topological
invariant we need Eq. (\ref{dN5}) to be satisfied. We rewrite this equation as
follows:
\begin{eqnarray}
\label{dN5_} 0 & = &  {\rm lim}_{p_5^2 \rightarrow \infty}\int_{\partial [{\cal
M}\otimes R]} \, {\bf tr} \left( \frac{1}{{\cal G}^{-1} + p_5\gamma^5}[\delta
{\cal G}^{-1}]\Bigl( \wedge\frac{1}{{\cal G}^{-1} + p_5\gamma^5} \,
  d  { \cal G}^{-1} \Bigr)^4  \right)
\end{eqnarray}
For any values of $p$ we get the limit $p^2_5\rightarrow \infty$ first. That's
why Eq. (\ref{dN5_}) is satisfied for the system of overlap fermions with the
interactions turned on and $\tilde{\cal N}_5$ also remains the topological
invariant.

\subsection{Index theorem for overlap fermions}

In the intermediate states at $m_0 = 2,4,6,8, m\ne 0$ there are no true
massless states. Using data of Table \ref{table4} one finds that this is in
accordance with Eq. (\ref{nfz1}) of the index theorem.
 However, there are the generalized unparticles with the Green
function given by Eq. (\ref{A0}) (see the definition in Section
\ref{SectIndTheor}). The number of generalized unparticles is related to the
jump in $\tilde{\cal N}_4$:
\begin{equation}
n_f^u = \Delta \tilde{\cal N}_4
\end{equation}
This relation can easily be checked and is also in accordance with theorem
\ref{indextheorem}.

In the intermediate state with $m = 0, m_0 \ne 0,2,4,5,8$ the generalized
unparticles are absent. Zeros of $\cal G$ remain zeros across the transition.
However, there are massless fermions. Their number is: $0$ for $m_0 < 0$, $1$
for $2 > m_0 > 0$, $5$ for $4
> m_0 > 2$, $11$ for $6 > m_0
> 4$, $15$ for $8 > m_0 > 6$, $16$ for $ m_0 > 8$. At the same time the number
of topologically protected massless fermions given by Eq. (\ref{nfz0}) is:
\begin{equation}
n_f = \frac{1}{2} \Delta \tilde{\cal N}_5
\end{equation}
This  is $0$ for $m_0 < 0$, $1$ for $2
> m_0
> 0$, $-3$ for $4
> m_0
> 2$, $3$ for $6
> m_0 > 4$, $1$ for $8 > m_0 > 6$, $0$ for $ m_0 > 8$. Therefore, except for
the conventional case $2 > m_0 >0$ there are massless fermions in the
intermediate states that are not protected by momentum space topology. When the
interaction with the gauge field is turned on some of them may annihilate each
other so that the total number of massless fermions is reduced without the
phase transition.

There are also mixed intermediate states $m = 0, m_0 = 2,4,6,8$, where both
massless fermions and generalized unparticles are present. The corresponding
transitions satisfy the conditions of  Theorem \ref{indextheorem2}. All pseudo
- poles of the Green function become true poles. The zeros of the Green
function may be transformed to the massless excitations across the transition
points at $m  = 2,4,6,8$. At the corresponding points the generalized
unparticles appear. Their number is equal to the jump of $\tilde{\cal N}_4$.
The corresponding values are listed in Table \ref{table4}.

\begin{table}
\begin{center}
\begin{small}
\begin{tabular}{|c|c|c|c|c|c|c|c|c|c|c|c|c|c|c|c|c|}
\hline
$m_0$ & $\tilde{\cal N}_4$ & $\tilde{\cal N}_5$   \\
\hline
$-m_0 > 0 $ & $0$ & $0$ \\
\hline
$-2 < -m_0 < 0$ & $1$ & ${\rm sign}\, m$  \\
\hline
$-4 < -m_0 < -2 $ & $-3$ & $-3\,{\rm sign}\, m$ \\
\hline
$-6 < -m_0 < -4$ & $3$ & $3\,{\rm sign}\, m$  \\
\hline
$-8 < -m_0 < -6 $ & $-1$ & $-{\rm sign}\, m$  \\
\hline
$ -m_0 < -8 $ & $0$  & $0$ \\
\hline
\end{tabular}
\end{small}
\end{center}
\caption{The values of topological invariants $\tilde{\cal N}_4$ and
$\tilde{\cal N}_5$ for free overlap fermions. } \label{table4}
\end{table}

\section{Conclusions and discussion}
\label{Conclusions}

In the present paper we discuss the topological invariants in case when
momentum space of the $4D$ model admits non - analytical exceptional points
(related to the generalized unparticles) and zeros of the Green function, the
physical meaning of which is still not well recognized (see, however,
\cite{Gurarie2011,EssinGurarie2011}). The topological invariants introduced in
\cite{VZ2012} are generalized to the case, when the Green function may contain
zeros and poles. The index theorem is formulated that relates the number of
generalized unparticles and massless fermions at the quantum phase transition
with the jumps in the topological invariants.

The topological invariants do not feel smooth changes of the model. Only a phase transition may lead to the change of the topological invariant. Therefore, if the free system (without gauge fields) and the interacting system (with gauge fields) belong to the same phase, then the values of the topological invariants are the same. So, we may calculate the topological invariant for the free fermions and it will be equal to the same value for the complicated interacting system that is related to the free system by a smooth transformation. The latter condition is important.  That’s why, for example, all presented constructions cannot be used directly for the Aoki phase as it is not connected smoothly to the noninteracrting system: we always pass a phase transition in order to jump into this phase.

In order to illustrate the properties of the mentioned constructions we
consider the lattice models with overlap fermions.  We found that the vacuum
states of lattice models with fully gapped fermions but with zeros in Green
function (insulating vacua) in 4D space-time are characterized by two
topological invariants, $\tilde{\cal N}_4$ and $\tilde{\cal N}_5$. They are
responsible for the number of generalized unparticles and gapless fermions
which appear at the topological transitions between the massive states with
different topological charges.  According to Theorem \ref{indextheorem} the
jump in $\tilde{\cal N}_4$ determines the total number of unparticles in the
intermediate state, while both the jumps in $\tilde{\cal N}_4$ and $\tilde{\cal
N}_5$ are related to the number of massless fermions in the intermediate state.

The mentioned properties of the lattice regularization with overlap fermions
may be related not only to the technique of the given particular
regularization. Namely, the continuum limit at $m = 0, m_0 = 2,4,6,8$ may be
taken seriously. In such limit a continuum theory appears that contains the
unparticle excitations. At the same time, the general properties of the quantum
phase transition with change of $\tilde{\cal N}_4$, $\tilde{\cal N}_5$ can be
applied to the relativistic field theories with fermions.  So, we may have the
new look at the high energy field theoretical models. The entireties, that are
new  for the high energy physics, appear. These are the zeros and the non -
analytical exceptional points of the Green function. We relate the latter
points to the generalized unparticles thus extending the definition of
unparticles given by Georgi \cite{unparticle}. In fact, there is the important
difference between the two concepts: Georgi unparticles considered in
\cite{unparticle,fermion_unparticle,fermion_unparticle2} have propagators  that
behave as $\sim p^{2d_U -4}$ with some $d_U$. Our generalized unparticles are
fermionic objects. Their "propagators" are non - analytic at $p=0$, the non -
analyticity is of the general form. However, in case of noninteracting overlap
fermions at $m = 0, m_0 = 2,4,6,8$ the excitations appear with the propagator
$\sim \frac{\gamma_a p^a}{|p|}$ that has the form of the propagator derived in
\cite{fermion_unparticle} with $d_U = 2$.

One may suppose that the form of this propagator might remain the same when the
interactions with the gauge fields are turned on. If so, nontrivial
conventional unparticles with $d_U = 2$ would emerge in the continuum limit of
lattice fermion systems with overlap fermions at the position of the phase
transition between the massless vacua with different values of the topological
invariant $\tilde{\cal N}_4$. Gauging of unparticles is achieved, therefore, in
an obvious way. The setup for such an investigation may include, for example,
the consideration of the system that consists of the doublet of Dirac fermions
interacting with the $SU(2)$ gauge field. The model is to be considered in
lattice regularization with overlap fermions. According to the results of the
present paper when the interaction with the gauge field is switched off, at the
very weak gauge coupling,  and at $m_0 \rightarrow 2, m \rightarrow 0$ the
lattice model contains two massless fermion excitations and $8$ species of
fermionic unparticles. Both these two kinds of excitations are expected to
survive in continuum limit. So, we have the field theoretical description of
the unparticles. When the interaction is turned on, the generalized unparticles
may appear instead. How the unparticle propagator has changed in this case is
to be a subject of the further investigation.

The author kindly acknowledges discussions with G.E.Volovik. Actually, this was
his idea to relate the excitations Eq. (\ref{A0}) with the notion of
unparticle. This work was partly supported by RFBR grant 11-02-01227, by Grant
for Leading Scientific Schools 679.2008.2, by the Federal Special-Purpose
Programme 'Cadres' of the Russian Ministry of Science and Education, by Federal
Special-Purpose Programme 07.514.12.4028.

\end{document}